
\input phyzzx
\input epsf
\overfullrule=0pt

\def\p{\partial}

\def\ga{\alpha}
\def\gb{\beta }
\def\ap{\alpha '}

\def\IN{\relax{\rm I\kern-.18em N}}

\null
\rightline {UTTG-10-94}
\rightline {May 1994}

\title{Thermal ensemble of string gas in a
magnetic field
\foot{Work supported in part by NSF grant
PHY 9009850 and R.~A.~Welch Foundation.}}

\author{Jorge G. Russo \foot{
Address after September 1, 1994: Theory Division, CERN, CH-1211
Geneva 23, Switzerland} }
\address {Theory Group, Department of Physics, University of
Texas\break
Austin, TX 78712}

\abstract
We study the thermal ensemble of a gas of free strings in presence
of a magnetic field. We find that the thermodynamical partition function
diverges when the magnetic field exceeds some critical value $B_{\rm cr}$,
which depends on the temperature. We argue that there is a first-order phase
transition with a large latent heat.
At the critical value an infinite number
of states -all states in a Regge trajectory- seem to become massless,
which may be an indication of recuperation of spontaneously
broken symmetries.

\vskip 1.5cm

\bigskip
\endpage

\Ref\hagedorn {R. Hagedorn, Nuovo Cim. Suppl. 3 (1965) 147.}

\Ref\nqcd{A. Polyakov, Phys. Lett. B72 (1978) 477;
L. Susskind, Phys. Rev. D20 (1979) 2610.}

\Ref\otros{B. Sathiapalan, Phys. Rev. D35 (1987) 3277;
Ya. Kogan , JETP Lett. 45 (1987) 709;
K. O'Brien and C.-I. Tan, Phys. Rev. D36 (1987) 1184.}

\Ref\atwi{J.J. Atick and E. Witten, Nucl. Phys. B310 (1988) 291.}

\Ref\amati{D. Amati, Ciafaloni and G. Veneziano,  Phys. Lett. B197 (1987)
81; Nucl. Phys. B403 (1993) 707.}

\Ref\grosmen{D. Gross and P.F. Mende, Nucl. Phys. B303 (1988) 407;
Phys. Lett. 197B (1987) 129.}

\Ref\gross{D. Gross, Phys. Rev. Lett. 60B (1988) 1229.}

\Ref\moore{G. Moore, {\it Symmetries of the bosonic string S-matrix},
Yale University preprint, YCTP-P19-93 (1993); YCTP-P1-94 (1994).}

\Ref\salam{A. Salam and J. Strathdee, Nucl. Phys. B90 (1975) 203.}

\Ref\nielsen{ N.K. Nielsen and P. Olesen, Nucl. Phys. B144 (1978) 376;
J. Ambj\o rn and P. Olesen, Nucl.Phys. B 315 (1989) 606; {\it ibid}
B330, (1990) 193.}

\Ref\porrati{ S. Ferrara and M. Porrati, Mod. Phis. Lett. A8 (1993) 2497.}

\Ref\russus{ 
J.~G.~Russo and L.~Susskind,
``Asymptotic level density in heterotic string theory and rotating black
holes,''
Nucl.\ Phys.\ B {\bf 437}, 611 (1995)
[hep-th/9405117].
}

\Ref\ferrara{E. Del Giudice, P. Di Vecchia and S. Fubini, Ann. Phys. 70
(1972) 378; K. A. Friedman and C. Rosenzweig, Nuovo Cimento 10A (1972) 53;
S. Matsuda and T. Saido, Phys.Lett. B43 (1973) 123; M. Ademollo {\it et al},
Nuovo Cimento A21 (1974) 77;
S. Ferrara, M. Porrati and V.L. Teledgi, Phys. Rev. D46
(1992) 3529.}

Among the mysteries clouding a fundamental formulation of
string theory, the Hagedorn transition is, probably, the most arcane.
It is found that the thermodynamical partition function of a free string
gas diverges at some finite temperature [\hagedorn ]. The analogy with
large $N$ QCD [\nqcd ] suggests that the
Hagedorn temperature may not be a limiting temperature but rather an
indication of another phase of the theory, perhaps where the description
of physics in terms of strings is inadequate. This interpretation is
supported by the results of refs. [\otros ], where it is shown that at
the Hagedorn transition a certain mode  becomes massless.
By studying the effective field theory near the Hagedorn temperature,
Atick and Witten [\atwi] argued that the transition should be first order
with a large latent heat, due to a genus-zero contribution to the free energy
above the Hagedorn temperature.

In quantum chromodynamics a clear evidence of partons appears
in high-energy scattering processes. High-energy scattering also provides a
way to recognize spontaneously broken symmetries.
In the case of string theory,
diverse studies in this direction were made in refs. [\amati -\moore ].
In particular, in ref. [\gross ] it was argued that in the
high-energy limit string scattering amplitudes obey an infinite number
of linear relations that are valid order by order in perturbation
theory. This suggests the existence of an enormous symmetry which is
restored at high energy.

In superconductivity, the restoration of the U(1) symmetry is achieved
either by increasing the temperature or by increasing the magnetic field.
This effect gives rise to the well-known Meissner curve separating the
superconducting phase from the normal phase in type I superconductors.
The analog of this phenomenon in the context of particle physics was explored
in ref. [\salam ], and more recently in ref. [\nielsen ],
 where it was argued that spontaneously broken symmetries by
the Higgs mechanism can be restored in the presence of a strong magnetic field.

In this paper we will further explore the Hagedorn transition by
considering a string gas in a magnetic field. We will argue that
a phenomenon analog to the case of superconductivity
takes place in string theories.
We will also argue that there are genus zero contributions to the free
energy above the critical magnetic field, and
obtain the critical $B-T$ curve.
In addition we will consider the heterotic string theory and find that
an infinite number of physical particles
become massless for approximately the same critical
value of the magnetic field. \foot {In the context
of (zero-temperature) open string theory this was first pointed out in
ref. [\porrati ], where it was also argued that this fact indicates a
phase transition with possible restoration of symmetries.}
We will find that this value of the magnetic field is precisely
the critical value beyond which the thermodynamical partition function
diverges.
This result supports the view that there should
exist an infinite-dimensional symmetry group governing string interactions,
and it may provide a clue on the organization of multiplets.

Let us first derive an asymptotic formula for the level density of states
with mass $M$ and angular momentum $J$ in the case of the bosonic open string
theory. This calculation was done in [\russus ] and we refer to this paper
for details. Here
we will present a simple, alternative derivation. We add to the world-sheet
Hamiltonian
a term containing the angular momentum in the $z$ direction with a
Lagrange multiplier,
$$
H=\sum_{n=1}^\infty \sum_{i=1}^{D-2}
\ga_{-n}^i\ga_{n}^i+\lambda J\ ,\ \ \
J=-i\sum_{n=1}^\infty {1\over n}\big( \ga_{-n}^1\ga_{n}^2
-\ga_{-n}^2\ga_{n}^1\big)\ .
\eqn\hamm
$$
 The Hamiltonian can be diagonalized by
$\ga_n^1=\sqrt{n/2} \big(a_n+b_n\big)\ ,\
\ga_n^2=-i\sqrt{n/2} \big(a_n-b_n\big)\ $.
One obtains for the partition function,
$Z=\tr \big[ e^{-\beta H}\big] $, the following expression:
$$
Z=\prod _{n=1}^\infty \bigg[ \big( 1-w ^n\big) ^{-D+4}
\big( 1-cw^n\big) ^{-1}\big( 1-{w ^n\over c} \big)^{-1}\bigg ]\ ,
\eqn\partitio
$$
where $w \equiv e^{-\beta} $ and $c\equiv e^{\beta \lambda}$.
Let us define $G=\log Z$ and consider
$$
{\p G\over\p \log c}=\big( c-c^{-1}\big)\sum_{k=1}^\infty
{w^k\over \big( 1-cw^k\big) \big( 1-c^{-1}w^k\big)}\ .
\eqn\cinco
$$
Inserting $c=e^{\gb \lambda }$ and taking the limit $\gb \to 0 $
we obtain
$$
{\p G\over\p \lambda }=2\lambda \sum_{k=1}^\infty {1\over k^2-\lambda^2}\ .
\eqn\seis
$$
The summation in eq. \seis\ can be performed explicitly. Indeed
$$\eqalign {
{\p G\over\p \lambda }&=- \sum_{k=1}^\infty \bigg(
{1\over k+ \lambda }-{1\over k-\lambda}\bigg)=\psi(1+\lambda) -
\psi(1-\lambda) \cr
&=\int_0^1 dx {x^\lambda-x^{-\lambda}\over x-1}={1\over\lambda}-
\pi {\rm cotg} (\pi\lambda ) \ .\cr }
\eqn\siete
$$
Thus
$$
G(\lambda )=\log {\lambda\over\sin (\pi\lambda )}+ G(0)\ .
\eqn\ocho
$$
The term $G(\lambda=0)$ can be obtained by writing $ Z$ in the following
way:
$$
Z(w, c)=\exp\big[ \sum_{m=1}^\infty {1\over m}\big( c^m+c^{-m}+D-4 \big)
{w^m\over (1-w^m)}\big]\ .
\eqn\masparti
$$
and setting $c=1$. The leading term as $\gb\to 0 $ is
$Z(w,0)=e^{a^2\over\gb }, \ a\equiv \pi \sqrt {(D-2)/ 6}$.
In this process subleading, power-like factors are neglected.
Thus we obtain
$$
Z(\gb ,\lambda )\cong {\rm const. } e^{a^2\over\gb }{\lambda\over
\sin (\pi \lambda )}\ .
\eqn\zetap
$$
This estimate is in agreement with the more accurate calculation
of ref. [\russus ].

By expanding  $Z$, $Z(w,k)=\sum_{n,J} d_{n,J}w ^n e^{ik J} ,\ k
=-i\beta\lambda $, then $d_{n,J}$ can be found by
$$
d_{n,J}={1\over 2\pi i}
\oint {dw\over w^{n+1}} \int_{-\infty} ^\infty {dk\over 2\pi} e^{-ik J}
\ Z(w,k )\ .
\eqn\ddnj
$$
where the contour goes over a small circle around $w=0$.
These integrals were carried out in ref. [\russus ], with the result
$$
d_{n,J} \cong {\rm const.} e^{(n+1)\gb+ a^2/\gb }
{1\over {\rm cosh}^2(\gb J/2) } \ ,\ \gb\equiv {a\over\sqrt{n+1-|J|}}\ .
\eqn\uno
$$
The constant can be chosen so that $d_{n,J}=1$ on the Regge trajectories
$J= \pm n$.

Let us consider a canonical ensemble of free string gas in the
presence of a magnetic field in the $z$ direction. We have
$$
Z(T,B)=\int_0^\infty dn \int_{-n}^n dJ d_{n,J} e^{-\gb E}\ ,
\eqn\therpar
$$
where
$$
E^2=m^2+2qB(l+1/2)-2JqB+O(B^2)\ ,\ \ \ap m^2=n-1
\eqn\energi
$$
$l$ represent the Landau level, and $q$ is the total Chan-Paton charge
of the open string. A sum over $l$ is also understood in
eq. \therpar . In eq. \energi\ we have used the fact that all
physical states in open string theory have gyromagnetic factor equal 2
[\ferrara , \russus ].
The terms of $O(B^2)$ have various origins. In particular,
a magnetic field generates non-trivial corrections to sigma-model
backgrounds starting from $O(B^2)$. Let us disregard for the moment $O(B^2)$
terms. From eq. \energi\ there seems to be a critical magnetic field at which
some states become tachyonic, as noted in ref. [\porrati ].
The first states to become tachyonic
are those with maximum spin, $J=n$, at a magnetic field
$$
qB_{\rm cr}\cong {1\over 2\ap-{1\over m^2}}\ .
\eqn\becriti
$$
The state on the Regge trajectory with $\ap m^2=1$ will become tachyonic
at $qB\cong 1/\ap $.
For large $m^2$ all the states on the first Regge trajectory will become
tachyonic at $qB_{\rm cr}\cong 1/2\ap $.
Massless states with spin 1 will be tachyonic for an infinitesimal
value of the magnetic field. This last effect may be separated,
as will be done in the case of the heterotic string theory.

The emergence of tachyons is a clear sign of instability.
It is energetically
more favorable to produce pairs with maximum spin aligned with the
magnetic field. As a result,  the partition function \therpar \ will
diverge for $B$ exceeding the critical value. This can be combined
with another well-known effect.
At zero magnetic field, the partition function diverges
if the temperature is above the Hagedorn temperature, which
for the bosonic open string theory is $T_H=1/4\pi\sqrt{\ap }$.
A critical curve $B-T$, analogous to the Meissner curve in superconductivity,
 can be obtained by inserting  eq. \uno\
into eq. \therpar \ and analysing the convergence properties.
For this purpose  power-like dependence can be neglected relative
to the exponential factors.
Let us write $J=\ga n$ and analyse the integrand of \therpar\ in the
region $-1<\ga <1 \ ,\ n\to\infty $. By using eqs. \energi\ and \uno\
we obtain the conditions
$$
\ga=0:\ \ d_{n,J} e^{-\gb E}\to 0 \iff T<T_H\ ,\ \ T_H=1/4\pi\sqrt{\ap }
$$
$$
\ga=1:\ \ d_{n,J} e^{-\gb E}\to 0 \iff B<B_{\rm cr}\ ,\ \ B_{\rm cr}=1/2\ap
$$
One can verify that these are necessary and sufficient conditions for
convergence in all the region $-1<\ga <1 \ ,\ n\to\infty $.

The phase diagram is displayed in Fig. 1.
Our analysis  only provides
a rough estimate. It is plausible that a more careful and systematic
inclusion of $O(B^2)$ and other effects will smooth out the critical curve.
In particular,
if the transition is first order one expects that the actual critical
temperature at zero field is below the Hagedorn temperature [\atwi ].
The derivative of $B$ in the coexistence curve ${dB_{\rm cr}\over dT}$
at $T=0$ should remain zero in a full treatment. There is
a thermodynamical reason for this. Indeed, on
the coexistence curve the chemical potentials of the phases must be equal,
and thus one has $S_n-S_s\prop {dB_{\rm cr}\over dT}$. But the third
law of thermodynamics implies that $S_n-S_s\to 0$ as $T\to 0$.

In any case, it is possible that the critical line is actually
not well defined,
since  the notion of temperature may cease to be valid at the Planck scale.

In superconductivity the critical magnetic field $B_{\rm cr}$ is related
thermodynamically to the difference of the Helmholtz free energy density of
the two phases. One has
$$
{B_{\rm cr} ^2\over 8\pi }={F_n(T)-F_s(T)\over V}\ .
$$
The curve $B_{\rm cr}(T)$ is quite well approximated by a parabolic law,
$B_{\rm cr}(T)\cong B_{\rm cr}(0)\big[1-\big(T/T_{\rm cr}\big)^2\big]$.

\vskip 15pt
\vbox{{\centerline{\epsfxsize=3.25truein \hskip 2cm \epsfbox{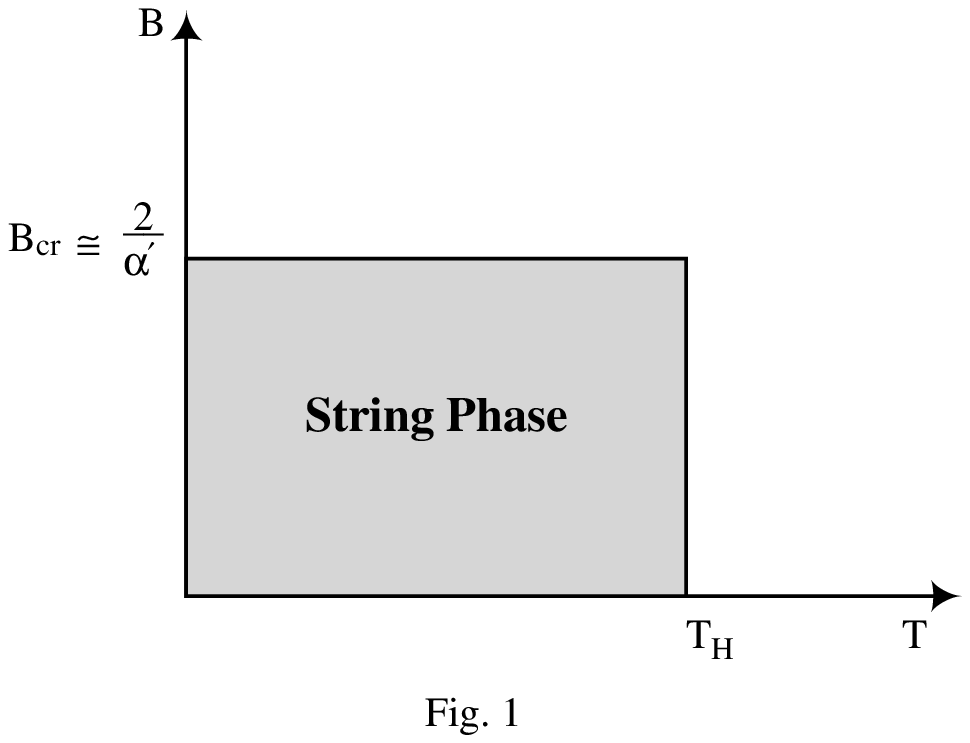}}
\vskip 12pt
{\noindent{\tenrm FIGURE 1. Phase diagram $B-T$.}}
\vskip 15pt}}

By using eq. \uno ,
in the case of the heterotic string theory one gets ($n_L\cong n_R\equiv n$)
$$
d_{n,J_L,J_R} \cong
{\rm const.} { e^{(n+1)(\gb_L+\gb_R)+ a^2_L/\gb_L+a^2_R/\gb_R }
\over
\cosh^2(\gb _LJ_L/2)\cosh^2(\gb_R J_R/2) } \ ,
\gb_L\equiv {a_L\over\sqrt{n+1-J}}\ ,\ \gb_R\equiv {a_R\over\sqrt{n+1-|J|}}\ ,
\eqn\dhetero
$$
where $a_L=2\pi $ and $a_R=\sqrt{2}\pi $ and $J_{L,R}$ represent
the left and right contribution to the angular momentum. Thus
$$
d_{n,J}= \int_ {-n}^n dJ_L d_{n,J_L,J-J_L} =\int_ {-n}^n
 dJ_R d_{n,J-J_R,J_R}\ .
\eqn\dtotal
$$

The energy formula can be obtained by using the gyromagnetic coupling
derived in ref. [\russus ]. One obtains
$$
E^2=m^2+2qB(l+{1\over 2})- 2J_RqB +O(B^2)\ ,\ \ \ap m^2/2\cong 2n ,\ n>>1\ .
\eqn\enerhet
$$
Again there appears to be a limiting or critical  value for the magnetic
field.
It may be convenient to go away from the self-dual point,
so that all the particles with gyromagnetic coupling have mass.
When
$$
qB_{\rm cr}\cong {2\over \ap-{2\over m^2}}\ ,
\eqn\campob
$$
all the states on the Regge trajectory with $J_R=n$ become massless.
Let us consider the partition function
$$
Z(T,B)=\int_0^\infty dn \int _{-n}^n dJ_L \int _{-n}^n dJ_R
d_{n,J_L,J_R}e^{-\gb E}
\eqn\queseyo
$$
The peculiar coupling of the magnetic field with the right contribution
to the angular momentum is characteristic of heterotic string theories
where the gauge quantum numbers solely arise from the left sector
[\russus ].
A similar analysis as  in the open string case leads to the following
convergence region:
$$
qB< 2/\ap \ ,\ \ T<T_H^{\rm het}\ ,
\eqn\hetcurv
$$
with $T_H^{\rm het}=1/\big(\sqrt{\ap }(a_L+a_R)\big)={1\over\pi\sqrt{\ap }}
\big(1-{1\over\sqrt {2}}\big)$.

A natural question is whether  the critical magnetic field is not an
artifact of having ignored $O(B^2)$ terms. An complete treatment including
all orders in the $\ap $ expansion (i.e. including all powers of $B$)
would be necessary in order to answer this question. Certainly it
would be important to verify the existence of a critical magnetic field
in some specific, exactly solvable example.
However, it should be stressed that there is no reason why higher
orders in $B^2$ should stabilize the vacuum. It does not occur in the case of
superconductivity and it does not
seem to occur in the case of the electroweak model [\salam, \nielsen ].
$O(B^2)$ corrections to
the geometry,  etc. can only lead  to further instabilities of different
nature, as e.g. gravitational collapse.

Another way to derive a $B-T$ curve  is by considering heterotic string
with a compactified (euclidean) time dimension. Let us consider
 the right-moving modes in the NS sector (the analysis at $B=0$ was
carried out in ref. [\atwi ]). In this case the energy is given by
$$
\ap E^2=-3 +{4 \ap\pi^2 n_m^2\over \gb ^2}+ {n_w^2\gb ^2\over 4\ap \pi
^2}-2J_R \ap qB +2N+2\tilde N +O(B^2)\ ,
\eqn\windi
$$
where $n_m$ and $n_w$ respectively denote quantized momentum and winding
number. At zero field the first state to become tachyonic is
that with $N=\tilde N=0$ and $n_w=\pm 1, n_m=\pm 1/2$. This has $J_R=0$
and hence it does not have gyromagnetic coupling. The curve is simply
a vertical line at the Hagedorn temperature. The first charged
state with $J_R\neq 0$ is given by $\tilde N=1$, $N=1/2$, $n_m=0$ and
$n_w=2$ ($n_w=1$ is excluded by GSO projection -for odd $n_w$ it is reversed
relative to the standard projection). One obtains the curve
$qB=1/(\ap \pi T)^2 +O(B^2)$. This curve lies entirely above
the coexistence curve \hetcurv \ and
so it is not very relevant; the phase transition already occurs
for smaller fields.
 Above this curve the thermodynamical
partition function, obtained by calculating genus $\geq 1$ contributions
in the finite temperature theory ($X^0=X^0+\gb $),
will develop another divergence in
virtue of the appearance of the new tachyon.

In superconductivity the transition at zero magnetic field at $T_{\rm cr}$
is second order, but in the presence of a magnetic field there is a
discontinuous change in the thermodynamical state
of the system with an associated latent heat, and the transition is of first
order.
By a similar analysis as in ref. [\atwi ], using the effective field theory
of a state of level $n$ on the Regge trajectory that becomes tachyonic,
 one obtains that the free energy for $B>B_{\rm cr}$ is given by
$$
F\sim- {\rm const.}{n^2\over g^2} (4/\ap  -2qB)^2\ ,
\eqn\free
$$
where $g$ is the string coupling. This represents a genus zero
contribution, just as it happens at zero field above the Hagedorn
temperature. A genus-zero contribution cannot arise on
simply connected Riemann surfaces. In large $N$ QCD,
above the deconfining transition, continuum Riemann surfaces have to be
replaced by Feynman diagrams.

In refs. [\atwi , \gross ] it was argued that
string theory  might describe the
spontaneously broken phase of a highly symmetric theory.
This view is supported by what we have found here: the partition function
diverges and an infinite number of
particles become massless at approximately the same value of the magnetic
field. This suggests that an enormous gauge symmetry is being
restored.
This symmetry would relate higher spin particles, somehow circumventing
the Coleman-Mandula theorem, which asserts that the maximum spin
of a conserved charge cannot exceed 1, but assumes that the number of
particles with masses below any given scale is  finite. This
hypothesis does not apply in the $B\to B_{\rm cr} $ limit,
where symmetries would be recuperated, since infinitely many
particles are becoming massless simultaneously.
An interesting problem, which could unravel symmetries, is deriving an
effective field theory for all
particles in the Regge trajectory in a sigma-model background $B$ near
$B_{\rm cr}$.

The author is grateful to L. Susskind for helpful discussions
and useful remarks. He also wishes to thank W. Fischler for very
valuable and stimulating conversations.

\refout
\vskip 2cm

\vfill\eject
\end